# Performance Loss Analysis and Design Space Optimization of Perovskite Solar Cells

Sumanshu Agarwal[1, 2] and Pradeep R. Nair[3]

[1]Department of Energy Science and Engineering, Indian Institute of Technology Bombay,

[2]Department of Electronics and Communication Engineering, Institute of Technical Education and Research, Siksha 'O' Anusandhan (Deemed to be University) Bhubaneswar,

[3]Department of Electrical Engineering, Indian Institute of Technology Bombay

sumanshu@iitb.ac.in, prnair@ee.iitb.ac.in

*Abstract* — While the performance enhancement witnessed in the field of perovskite solar cells over the recent years has been impressive, it is now evident that further optimization beyond the existing literature would require detailed analysis of various loss mechanisms. Here we address the same through detailed numerical simulations of optical and electrical characteristics. We quantify the various losses like optical losses (5-6%), recombination losses (3-4%), and resistive losses against the Auger limited practical efficiency limits. Moreover, we illustrate the schemes that result in reduction of these losses and eventual increase in efficiency. In addition, we extend the analyses to identify the optimum thickness of perovskite and the factors affecting the optimum thickness have been discussed in detail.

# I. Introduction

Solar photovoltaics (PV) is regarded as one of the most promising renewable energy technologies and is expected to play an important role towards meeting the ever-increasing global energy needs. Perovskite solar cell (PSC) is a new class of solar cell in this field and has already achieved >22% efficiency[1]–[3]. Ambipolar nature of carrier transport with diffusion lengths being more than 1 μm[4]–[7] along with very large extinction coefficient[8]–[10] makes perovskites an ideal material for photovoltaics. Moreover, perovskite can be deposited using relatively simple techniques like spin coating[11]–[14]. There have been many exciting developments in the field of PSCs like, fabrication of high efficiency devices[1], [15]–[17], bandgap tuning[18], theoretical investigations on the physical mechanisms[19]–[21] that govern the performance, design and fabrication of tandem solar cells based on perovskites[22]–[24], etc. Indeed, detailed analyses of different detrimental aspects for PSCs is a topic of immense interest to the community for further design space optimization. In this context, here we provide a comprehensive study of optical and electrical losses in planar perovskite solar cells.

We note that there have been a few recent reports on loss analysis of perovskite solar cells – although with some limitations. For example, loss analysis reported ref. [21], lacks in identification of optimal thickness. On the other hand, thickness optimization reported in the literature[19], [25], lacks in quantification of losses due to the different mechanisms. Specifically, carrier recombination in the bulk of perovskite was neglected by Sun et al.[19], which is not an entirely valid assumption. Further, they predict that the efficiency is invariant with active layer thickness beyond a certain value for p-i-n structures while, on the contrary, there are experimental results which indicate that efficiency peaks for certain active layer thickness [26]–[28]. Minemoto et al.[25] performed thickness optimization calculations considering only SRH recombination in the active layer along with doped layers. In contrast, here we compute the optimal thickness by combining both optical and electrical modeling with explicit consideration of all three types of bulk recombination mechanisms (Radiative, Auger, and SRH) in the active layer. Accordingly, this enables us to provide better estimates for the optimal thickness of perovskite layers.

Schematic of a typical PSC is shown in Figure 1 where perovskite is active layer and acts as light harvester. Photo-generated electrons and holes in perovskite are transported to respective electrode selectively via electron transport layer (ETL) and hole transport layer (HTL). High band offsets at ETL/perovskite and perovskite/HTL interfaces[29] negate the possibility of non-selective carrier collection. Though PSCs have good carrier collection efficiency at short circuit conditions[30] and relatively high $V_{OC}$, these cells suffer with low $J_{SC}$ and low $FF$ resulting in practical efficiency less than theoretical limit (~29.5%). Analyses show that these cells can have $J_{SC}$ ~27.3 mA/cm$^2$ and $FF$ as high as 91% under 1 sun illumination but the current state of the art devices have $J_{SC}$ ~22 mA/cm$^2$ and $FF$ ~70%. Our detailed optoelectronic modeling of these cells indicate that a significant portion of solar spectrum with energy more than the band gap energy of active layer does not contribute to generation of charge carriers in the active layer and hence result in low $J_{SC}$, while resistive effect of device causes loss in $FF$. Recombination of charge carriers in the active layer is mainly reflected in the loss in $V_{OC}$ of the cell. We identify that though $J_{SC}$ can be increased by increasing the thickness of the active layer, $V_{OC}$ and $FF$ decrease with the thickness and result in an optimum thickness of perovskite. We also identify the dependency of optimum thickness on different performance improvement schemes such as contact layer doping, improved lifetime of charge carriers in the active layer, surface texturing/grating to enhance the $J_{SC}$, etc.

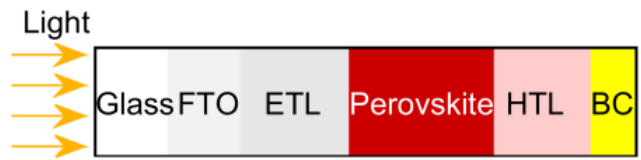

Fig 1. Schematic of a typical Perovskite solar cell (PSC). Here BC stands for back contact. Light incident on the device encounters reflections from air/glass, glass/FTO, FTO/ETL, ETL/perovskite, perovskite/HTL, HTL/BC, and BC/air interfaces.

## II. Model System

A typical PSC consists of a transparent conducting oxide (TCO) coated glass as substrate over which ETL, Perovskite, HTL, and back contacts (BC) are deposited (see fig 1). Here we consider FTO as TCO, TiO$_2$ as ETL, Spiro-MeOTAD as HTL, and MAPbX$_3$ (X= I, Cl) as active layer. The essential physical mechanisms to consider in any solar cell are – (a) the carrier generation rate due to optical absorption and (b) the bias dependent carrier collection at appropriate electrodes in the presence of various recombination mechanisms. To address (a), we use the transfer matrix methodology, described by Pettersson et al.[31] and assume that each absorbed photon in active layer result in generation of one e-h pair. We address the carrier collection through self-consistent solution of continuity and Poisson's equations using calibrated estimates for carrier generation rates[32], [33]. Details of equations used for optical and electrical calculations are provided in supplementary material. Material parameters for simulations are adopted from literature[34].

## III. Losses in perovskite solar cell

Literature indicates that Auger recombination coefficient for perovskite is of the order of $1 \times 10^{-29}$ cm$^6$/s. The JV characteristics using this estimate for Auger recombination coefficient and limiting rate of radiative recombination[35], [36] are shown in fig 2. The corresponding limits for performance parameters are $\eta = 29.5\%$, $FF = 91\%$, $J_{SC} = 27.3$ mA/cm$^2$, and $V_{OC} = 1.16$ V, where the efficiency of the cell defined in terms of $J_{SC}$, $V_{OC}$, and $FF$ is as follows:

$$\eta = \frac{J_{SC} V_{OC} FF}{P_{in}} \times 100. \quad (1)$$

We observe that experimental devices lag significantly behind these limits. For example, experimental JV characteristics of a planar PSC[4] are also plotted in the figure 2 and compared with the Auger limit JV curve. We find that Auger limited curve deviates a lot from the experimental JV curve. This deviation is observed in all the characteristic parameters of the solar cell viz. $J_{SC}$, $V_{OC}$, $FF$, and efficiency. From eq. 1, the deviation in the efficiency ($\Delta\eta$) is given as

$$\frac{\Delta\eta}{\eta} = 1 - \left(1 - \frac{\Delta J_{SC}}{J_{SC}}\right)\left(1 - \frac{\Delta V_{OC}}{V_{OC}}\right)\left(1 - \frac{\Delta FF}{FF}\right). \quad (2)$$

The relative contribution of each of the performance metrics towards the efficiency loss can be identified by retaining only the first order terms in eq. (2). Accordingly, the loss in efficiency could be given by

$$\frac{\Delta\eta}{\eta} \simeq \frac{\Delta J_{SC}}{J_{SC}} + \frac{\Delta V_{OC}}{V_{OC}} + \frac{\Delta FF}{FF}. \quad (3)$$

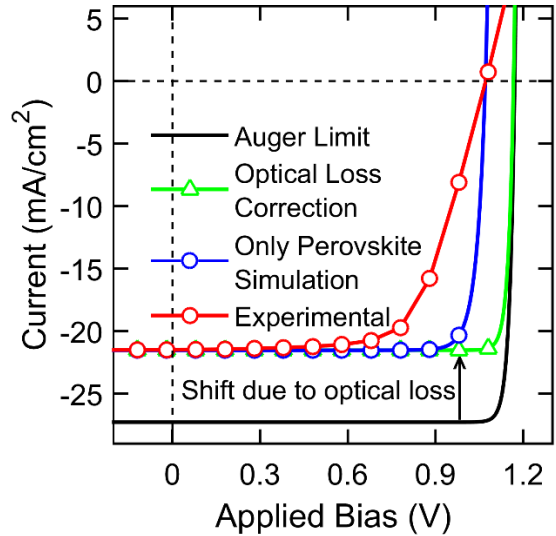

Fig 2. Light JV characteristics of PSC. Light JV characteristics for experimental device [4] is compared with the corresponding JV characteristics under Auger limit, Auger limit with optical loss, and device without contact layers (ETL and HTL) but the selective contacts i.e. only perovskite.

In the reported experimental data[4], loss in $J_{SC}$, $V_{OC}$, and $FF$ as compared to Auger limit is ~5.8 mA/cm², ~0.1 V, and ~24% respectively (here, it should be noted that though the state of the art efficiency of perovskite solar cell is 22.1%[3], [37], most of the devices exhibit ~15% efficiency. Given this, we compare our results with the experimental data reported by Liu et al.[4]). Hence, overall loss in efficiency as calculated from equation 3 is ~16.4% while the actual deviation is ~14%. Though the calculated loss is different from the actual loss (which is due to the assumptions involved in eq. 3, while eq. 2 is accurate), this calculation indicates that loss in the performance is mainly due to $FF$ (~7.8%) followed by $J_{SC}$ (~6.3%) and $V_{OC}$ (~2.3%) (numbers in the parentheses are absolute values). Since loss in the generation free charge carriers and inefficient collection of photogenerated charge carriers result in loss of performance parameters, we analyze them using optical and electrical simulations of the devices in the following sections.

## IV. Optical losses

For the present study we assume that perovskite has band gap 1.55 eV[38]. We find that most of the reported devices with high efficiency show $J_{SC}$~22 mA/cm² compared to Auger limit $J_{SC} = 27.3$ mA/cm². To calculate the loss in $J_{SC}$ we first calculate photocarrier generation rate in the device using transfer matrix method (TMM)[31] (see supplementary material for details). The thicknesses of TiO$_2$, perovskite, and spiro, we use, are 225 nm, 300 nm, and 200 nm, respectively[39].

We find that transmittance of device is zero but approximately 14% incident light is lost in reflection and 7% is absorbed by other layers. Hence, only 79% of available sunlight with energy more than perovskite bandgap energy generates electron-hole pairs in the device. Figure 3a shows the reflectance of the device and absorptance of different layers for different wavelength. Due to large refractive index of TiO$_2$ as compared to perovskite at shorter wavelength, our calculations indicate high reflectance in the device at short wavelengths (280-320 nm). While we observe non-zero reflectance for complete spectrum, parasitic absorption is present only for short wavelengths (300-400 nm, in TiO$_2$) and long wavelengths (600-800 nm, in Spiro and Ag). The absorptance of active layer is very high in the wavelength range 400-600 nm, and this accounts for most of the current density in the device. Figure 3b shows the amount of photons reflected/absorbed by device/different layers under one sun illumination (AM 1.5 spectrum).

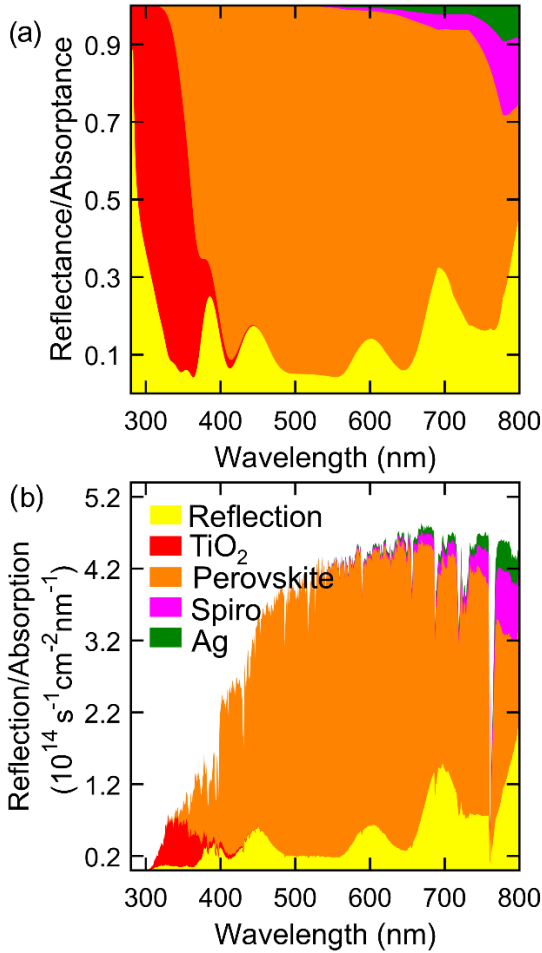

Fig 3. Reflectance from the device and absorptance of different layers. a) shows the normalized value of reflectance for the device and absorptance by different layers while b) shows the photon count reflected by the device and absorbed by different layers.

On comparison of figures 3a and 3b, we find that though absorptance of $TiO_2$ at shorter wavelength is higher than absorptance of spiro or Ag at longer wavelength, the actual loss in photogeneration caused by parasitic absorption in spiro is more as compared to absorption in $TiO_2$. This is due to low photon count in solar spectrum around 300-400 nm wavelengths as compared to photon count at higher wavelengths (600-800 nm). It is clear from fig 3b that major loss in optical generation of charge carriers in active layer is caused by reflection from the device followed by parasitic absorption in spiro. Note that FTO and glass have zero extinction coefficient for the wavelength range of interest and therefore they do not cause any parasitic absorption loss in the device. The integrated values of photons reflected by the device or absorbed in different layers are given in table 1. This table clearly indicates that only 78.9% photons available in 280-800 nm range contribute to short-circuit current density while 21.1% are lost which consists of 14.2% reflection loss and 6.9% parasitic absorption loss. Our results broadly agrees with the results obtained through simulations of Maxwell's equations[21]. This analysis shows that for high $J_{SC}$ devices reflection losses are of paramount interest to be addressed.

Table 1: optical losses in different layers of perovskite cell

|  | **Reflected** | **Absorbed in** | | | |
|---|---|---|---|---|---|
|  |  | **TiO$_2$** | **Perovskite** | **Spiro** | **Ag** |
| **Photons (cm$^{-2}$s$^{-1}$)** | $2.4\times10^{16}$ | $3.5\times10^{15}$ | $1.3\times10^{17}$ | $5.4\times10^{15}$ | $2.9\times10^{15}$ |
| **Current (mA/cm$^2$)** | 3.8 | 0.6 | 21.5 | 0.9 | 0.5 |
| **%** | 14.2 | 2.0 | 78.9 | 3.2 | 1.7 |

For further insights, we analyze two distinct aspects of the reflection spectrum, viz., a) reflection due to the interfaces encountered by the light before entering in perovskite and b) reflection due to the interfaces encountered by the light after passing through perovskite. Since reflection losses from the device consist of reflections from all the interfaces in the device (including reflections from the interfaces encountered by the light after passing through perovskite) therefore, to calculate (a), we calculated reflectance spectra with increasing the thickness of perovskite. Higher thickness of perovskite lead to absorption of almost all the photons infiltrating in the perovskite and hence, component (b) becomes negligible. Such calculations show a saturation in reflection spectrum for perovskite thickness > 2μm (see fig S1 for details) and the saturated value of reflectance is ~9.5% under 1 sun illumination. This calculation is important to analyze the performance improvement schemes that reduce only component (a) of the reflectance spectrum.

To further analyze the losses, we calculated single pass reflectance for all the interfaces and transmittance for ETL, HTL, and perovskite. The corresponding results are plotted in figure 4. These calculations indicate that Glass/Air interface contributes maximum (4%) to reflection losses from the interfaces before perovskite, which is followed by reflection from FTO/TiO$_2$ and Glass/FTO interfaces. Major portion of reflections occurring from interfaces that are encountered before perovskite is lost either as parasitic absorption by ETL or reflection into the air. We observe that while Ag at the backside helps perovskite to absorb photons by reflecting back the light transmitted from perovskite (fig 4a), a significant amount of the reflected light escape to air as the refractive indices of TiO$_2$ and perovskite are similar for long wavelengths. Hence, high transmittance of perovskite for long wavelength (fig 4b) along with poor reflectance of perovskite/TiO$_2$ interface (eq S1 and fig 4a) result in high reflection losses (at large wavelengths). Accordingly, the reflection losses for higher wavelengths can be reduced by increasing the effective path length of the light in the perovskite or by using the perovskite with high extinction coefficient (result in reduced transmittance from perovskite, see fig S2).

## V. Recombination Losses

The other losses in the cell are due to electronic processes. The major components of these losses are recombination and resistive losses. While carrier recombination in perovskite and perovskite/transport layer interface affects the performance, carrier selective contact layers negate the possibility of recombination in the bulk of the transport layers or at the contact/transport layer interface. As such, we consider three types of bulk recombination mechanisms: a) trap assisted SRH, b) radiative, and c) Auger. A generalized equation for total rate of recombination ($R$) can be given by

$$R = \frac{np - n_i^2}{\tau_e(p+p_1) + \tau_h(n+n_1)} + B(np - n_i^2) + (A_n n + A_p p)(np - n_i^2), \quad (4)$$

where first term represents trap assisted SRH recombination, second term gives the amount of radiative recombination and third term denotes the Auger recombination. We neglect the effect of interface trap assisted recombination; however, the same can be coupled with the bulk SRH recombination using an effective lifetime. Moreover, the effect of interface trap assisted recombination at HTL/perovskite interface is shown to be negligible[40].

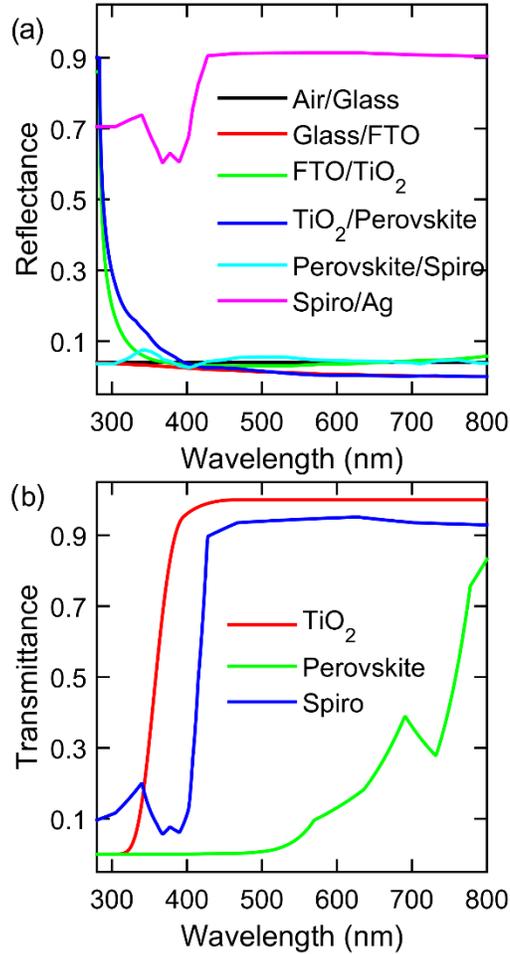

Fig 4. Reflectance from each interface and transmittance from different layers. a) shows the normalized value of reflectance for different interfaces in the device while b) shows the normalized value of transmittance from different layers. Glass and FTO have zero extinction coefficients and a large portion of light from spiro/Ag interface is reflected therefore transmittance of Glass, FTO, and Ag are not considered.

To differentiate between the losses due to active layer and contact layers, we performed device simulation of PSC in the absence of contact layers but the ideal selective contacts. Here ideal selective contacts mean electron (hole) recombination velocity is 0 while hole (electron) recombination velocity is infinite at cathode (anode). This simulation captures all the losses in the device caused by the perovskite layer. We used calibrated parameters for perovskite layer reported in the literature[34]. We have used uniform optical generation rate in perovskite layer ($G = 4.5 \times 10^{21}$ cm$^{-3}$-s$^{-1}$, obtained from optical calculations discussed above). We find that our device simulations with ideal carrier selective contacts in the absence of contact layers show perfect match in terms of $J_{SC}$ and $V_{OC}$ of reported experimental device with contact layers (see fig 2). A series resistance can anticipate the difference in the $FF$ in the modeled device and the experimental device. As such, series resistance correction of 9 Ωcm$^2$ to the JV characteristics of modeled device, results in experimental JV characteristics (see fig S3a). It indicates that experimental devices suffer with high series resistance offered by the presence of contact layers and results in degraded $FF$ and efficiency. Further, we observe that $J_{SC}$ loss ($\Delta J_{SC} = -5.8$ mA/cm$^2$) and $V_{OC}$ loss ($\Delta V_{OC} = -0.09$ V) corrections in auger limited JV curve result in the JV characteristics which are very similar to the device with ideal selective contact with small difference in $FF$ (see fig S3b). This difference in $FF$ is due to excessive recombination in simulated device at MPP, which is not incorporated by shifting current/voltage by $\Delta J_{SC}$ and $\Delta V_{OC}$.

## VI. Thickness Optimization and Performance Improvement Schemes

Different loss mechanisms discussed so far can be used to engineer a device with better performance. From previous discussions, it is intuitive that this can be done in two ways – a) reducing optical losses and b) reducing electrical losses. While (a) mainly contributes to increase in short circuit current density, (b) contributes to enhancement in $V_{OC}$ and $FF$ of the device. We performed detailed JV characteristics simulation by varying the thickness of active layer from 200 nm to 1200 nm and corresponding results are shown in figure 5. We used uniform optical generation in active layer as obtained from TMM method. Due to reduction in the reflection losses at higher wavelength with increase in the thickness of perovskite we observe increased generation rate and hence the $J_{SC}$. $J_{SC}$ vs thickness trend under the assumption of 100% $IQE$ is also shown in figure 5 (open diamonds without connectors). We observe saturation in terms of $J_{SC}$ at ~24 mA/cm$^2$ in this case.

Our results indicate that, due to reduction in electric field with increase in the thickness of active layer, $IQE$ is less than 100% for large thicknesses (see fig 5, actual $J_{SC}$ is less than expected $J_{SC}$ under the assumption of $IQE = 100\%$). Reduction in electric field result in low carrier collection efficiency[41], which results in reduction of $FF$ too (as shown in fig 5). Further, increase in the effective recombination due to increase in the thickness of active layer result in reduction of $V_{OC}$ (see figure 5). We also find that there is a maxima for efficiency and hence an optimum thickness.

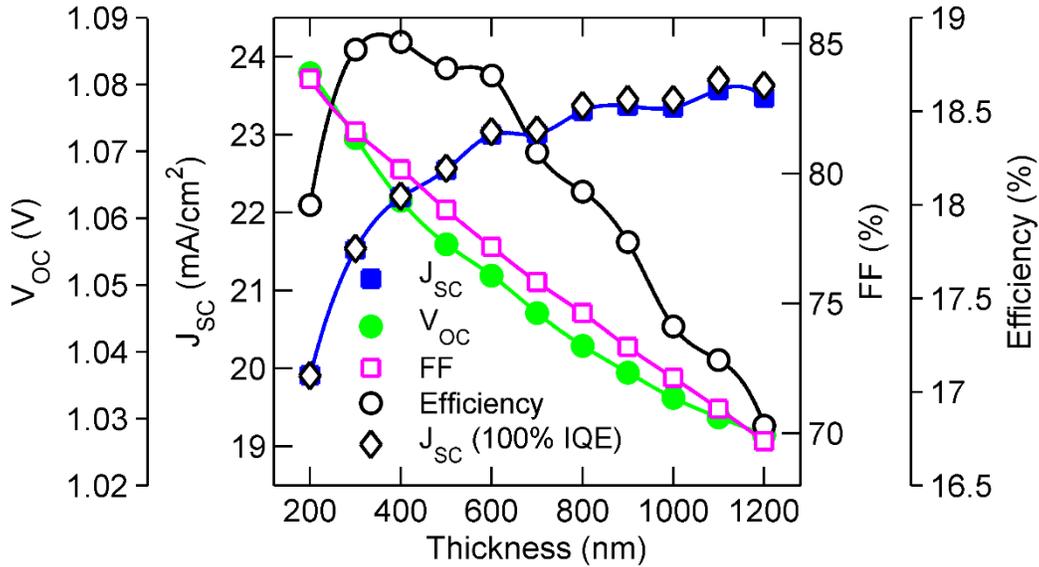

Fig 5. Effect of thickness on the performance of perovskite solar cells. Here we consider doped contact layers (doping density is 1×10$^{17}$ cm$^{-3}$) while the results for intrinsic contact layers are shown in fig S4. Open diamonds without connectors indicate the values of $J_{SC}$ (mA/cm$^2$) under 100% $IQE$ assumption.

Figure 5 has interesting implications in predicting the trends for optimal thickness. For example if a performance improvement scheme results in increase in $J_{SC}$ without affecting $V_{OC}$ and $FF$, then that scheme should result in decrease in optimum thickness (as the $J_{SC}$ vs thickness curve would move to higher values). Similarly, a scheme resulting in improvement in performance through $V_{OC}$ or $FF$ should show increase in optimum thickness. Accordingly, increase in the doping of contact layer show increase in optimum thickness (it improves the performance mainly through $FF$) and techniques like front surface texturing/grating show decrease in optimum thickness (performance improvement through $J_{SC}$). In addition, if the quality of the perovskite film is improved to reduce the rate of recombination in the bulk of the active layer then we observe increase in the optimum thickness (this scheme result in performance improvement through $V_{OC}$ and $FF$).

As expected, we observe high $FF$ in the devices with doped contact layers than devices with intrinsic contact layers (compare Figs. 5 and S4). Higher $FF$ of the devices with doped contact layers, due to – a) higher electric field in active layers (results in high collection efficiency[41]) and b) reduction in series resistance offered by contact layers, result in higher optimal efficiency. As predicted in previous paragraph, here the optimal thickness is more than the device with intrinsic contact layers.

Further, it is observed that the optimal thickness and efficiency are functions of rate of recombination of charge carriers in the active layer. We have plotted optimal thickness vs SRH lifetime in figure 6. As predicted earlier, we find that optimal thickness and efficiency increases with increase in SRH lifetime but after certain increase in SRH lifetime there is saturation in efficiency as well as optimal thickness of perovskite. Increase in performance is due to increase in effective lifetime which results in improvement in $V_{OC}$ and $FF$ [41], [42]. After certain increase in SRH lifetime, effective lifetime is controlled by radiative and Auger recombination and we do not observe any improvement further. For reduced radiative recombination rate (i.e. when $B \sim 10^{-13}$ cm$^3$s$^{-1}$, fundamental limit[34]) we observe further increase in optimal efficiency and optimal thickness. We find that increase in effective lifetime from basecase value (i.e. SRH lifetime=2.73 μs, $B = 3 \times 10^{-11}$ cm$^3$s$^{-1}$, and $A = 1 \times 10^{-30}$ cm$^3$s$^{-1}$) does not improve the $J_{SC}$, because collection efficiency at $J_{SC}$ is 100% for basecase (see fig 5).

In addition to this, we find that optimal thickness and efficiency are dependent on surface texturing too. Literature indicates that texturing of the front side reduces the reflection losses[43], [44] and hence result in higher $J_{SC}$. As discussed earlier, the reflection losses from all the interfaces before perovskite is ~9.5% (see discussion on optical loss and fig. S1). Accordingly 0% reflection losses (due to texturing) from all interfaces before perovskite should increase $J_{SC}$ by ~2.6 mA/cm$^2$. In this case, we find that optimum efficiency for device with base case SRH lifetime along with $B = 3 \times 10^{-13}$ cm$^3$s$^{-1}$ is ~21.5%. We also observe reduction in optimal thickness for basecase SRH lifetime as expected from fig 5. We also find improvement in performance on increasing the perovskite layer mobility (as reported elsewhere[45], mainly in terms of $FF$ due to increase in collection length[41], [42]) or extinction coefficient (lead to increased generation rate, see fig S2).

Figure 6a suggests that reported efficiency of 21.6% [1] on mesoporous structure could be due to improved –a) effective lifetime, b) optical generation, or c) proper doping of the contact layers. It is reported that the device show high $EQE$ ~90% (resulting in $J_{SC} = 22.7$ mA/cm$^2$), indicating higher rate of generation of charge carriers. Additionally, less nonradiative recombination in the active layer has been reported along with the doping of the contact layers. The combined effect of all these factors results in higher performance.

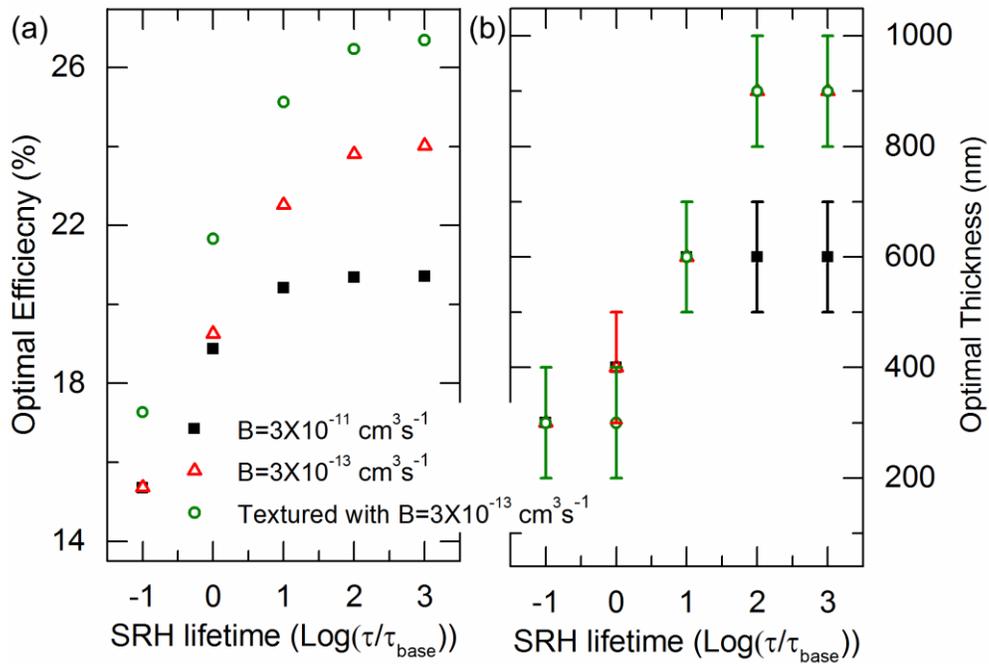

Fig 6. Variation in optimal performance of perovskite solar cells with carrier lifetime and photogeneration rate. (a) Optimal efficiency of the device vs SRH lifetime of perovskite. The curves for different B and generation rates are also plotted. Corresponding optimal thicknesses are plotted in (b).

Our analysis also indicate that for a given thickness of active layer, resistive losses through contact layers can be reduced by using the high mobility contact layers. Since the resistance of a semiconductor is inversely proportional to the product of mobility of charge carriers in the semiconductor and carrier density[32], any layer in the device with high mobility or high carrier concentration offers less resistive path. The high capacitance of $TiO_2$ due to its high dielectric constant along with low barrier height between conduction band of ETL and front contact (~0.2 eV) result in high carrier concentration in $TiO_2$ as compared to other layers. Therefore, resistance offered by $TiO_2$ layer is very low and we do not expect much improvement in efficiency by increasing its mobility. In contrast, low density of charge carriers in HTL results in high dependency of its resistance (hence the efficiency) on the mobility. Our simulations results (fig 7) also indicate that increase in HTL mobility as compared to $TiO_2$ mobility has more effect on the performance of the device and especially in terms of $FF$. Furthermore, increase in TiO$_2$ mobility has minimal effect on the efficiency of the cell as expected. Fig 7 shows that performance of the cell can be enhanced even with poor mobility TiO$_2$ substrate if mobility of HTL is sufficiently high. However, if a low dielectric material is used as ETL, then this layer also contribute to series resistance and corresponding measures need to be taken to reduce the resistance of the contact layers.

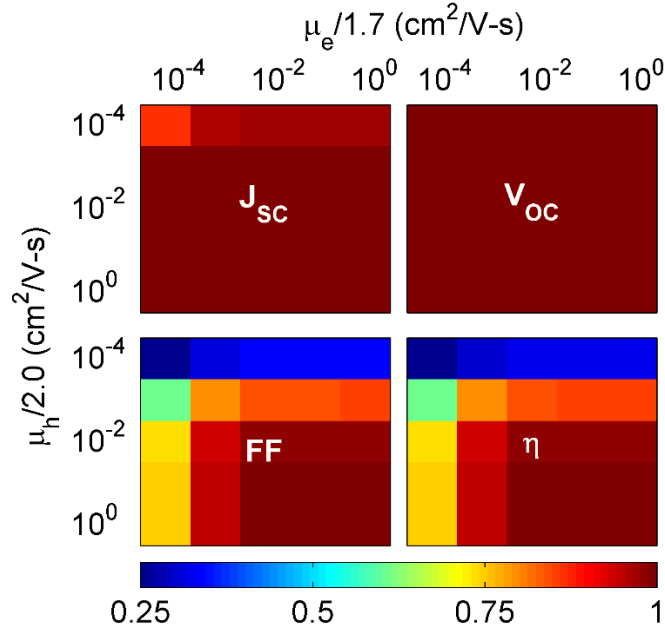

Fig 7. Performance trend of the device with variation in contact layers mobility. Nominal variation in $J_{SC}$ and $V_{OC}$ is observed while there are large variations in $FF$ and efficiency. The simulations were performed using base case parameters of perovskite. The normalization factors for $J_{SC}$, $V_{OC}$, $FF$, and $\eta$ are 21.52 mA/cm$^2$, 1.073 V, 80.1% and 18.47%, respectively.

### VII. Conclusions

To summarize, here we have identified different losses in the perovskite solar cells. For calibrated planar perovskite solar cell, we found that ~20% loss of optical generation along with resistive losses in the device and high recombination rate in active layer result in total ~9% (absolute) loss in efficiency which leads to ~20.5% efficiency of the cell in contrast to Auger limit of 29.5%. The result is very close to simulation result (20% efficiency) for device with ideal selective contacts. Further, we found the optimal thickness of perovskite for perovskite solar cells in planar architecture with optimal efficiency being ~20% for calibrated perovskite, which could further be improved to >25% by improving the material properties of the perovskite along with light trapping techniques. These results are in broad agreement with the experimental results of the literature where efficiency improvement has been observed due to light trapping and material parameter improvement.

## Acknowledgement

This paper is based upon work supported under the US-India Partnership to Advance Clean Energy-Research (PACE-R) for the Solar Energy Research Institute for India and the United States (SERIIUS), funded jointly by the U.S. Department of Energy (Office of Science, Office of Basic Energy Sciences, and Energy Efficiency and Renewable Energy, Solar Energy Technology Program, under Subcontract DE-AC36-08GO28308 to the National Renewable Energy Laboratory, Golden, Colorado) and the Government of India, through the Department of Science and Technology under Subcontract IUSSTF/JCERDC-SERIIUS/2012 dated 22$^{nd}$ November 2012. The authors also acknowledge the Center of Excellence in Nanoelectronics (CEN) and National Center for Photovoltaic Research and Education (NCPRE), IIT Bombay, for computational facilities.

Supplementary material for "Performance Loss Analysis and Design Space Optimization of Perovskite Solar Cells" can be provided upon request